\documentclass{sigchi}

\toappear{\scriptsize Permission to make digital or hard copies of all or part of this work for personal or classroom use is granted without fee provided that copies are not made or distributed for profit or commercial advantage and that copies bear this notice and the full citation on the first page. Copyrights for components of this work owned by others than ACM must be honored. Abstracting with credit is permitted. To copy otherwise, or republish, to post on servers or to redistribute to lists, requires prior specific permission and/or a fee. Request permissions from permissions@acm.org. \\
{\emph{CSCW 2015}}, March 14--18, 2015, Vancouver, BC, Canada. \\
Copyright \copyright~2015 ACM 978-1-4503-2922-4/15/03\ ...\$15.00. \\
http://dx.doi.org/10.1145/2675133.2675151
}

\usepackage{graphicx} % for EPS, load graphicx instead
\usepackage{times}    % comment if you want LaTeX's default font
\usepackage[hyphens]{url}      % llt: nicely formatted URLs
\usepackage{booktabs} % good tables

\makeatletter
\def\url@leostyle{%
  \@ifundefined{selectfont}{\def\UrlFont{\sf}}{\def\UrlFont{\small\bf\ttfamily}}}
\makeatother
\def\pprw{8.5in}
\def\pprh{11in}

\usepackage[pdftex]{hyperref}
\hypersetup{
pdftitle={SIGCHI Conference Proceedings Format},
pdfauthor={LaTeX},
pdfkeywords={SIGCHI, proceedings, archival format},
bookmarksnumbered,
pdfstartview={FitH},
colorlinks,
citecolor=black,
filecolor=black,
linkcolor=black,
urlcolor=black,
breaklinks=true,
}

\newcommand\tabhead[1]{\small\textbf{#1}}
\newcommand{\ownshown}{\textit{Own-Shown}}
\newcommand{\othershown}{\textit{Other-Shown}}
\newcommand{\bothshown}{\textit{Both-Shown}}
\newcommand{\allcond}{\textit{All users}}

\newcommand{\ownalgo}{\textit{Own Algorithm}}
\newcommand{\otheralgo}{\textit{Other Algorithm}}

\begin{document}

\title{Studying and Modeling the Connection between People's Preferences and Content Sharing}

\numberofauthors{2} 
\author{
\alignauthor
{Amit Sharma}\\
     \affaddr{Dept. of Computer Science}\\
       \affaddr{Cornell University}\\
      \affaddr{Ithaca, NY 14853 USA}\\
    \email{asharma@cs.cornell.edu}
% 2nd. author
\alignauthor
Dan Cosley\\
     \affaddr{Information Science}\\
   \affaddr{Cornell University}\\
 \affaddr{Ithaca, NY 14853 USA}\\
\email{danco@cs.cornell.edu}
}

% Teaser figure can go here
%\teaser{
%  \centering
%  \includegraphics{Figure1}
%  \caption{Teaser Image}
%  \label{fig:teaser}
%}

\maketitle

\begin{abstract}
People regularly share items using online social media. However, people's decisions around sharing---who shares what to whom and why---are not well understood. We present a user study involving 87 pairs of Facebook users to understand how people make their sharing decisions. We find that even when sharing to a specific individual, people's own preference for an item (individuation) dominates over the recipient's preferences (altruism). People's open-ended responses about how they share, however, indicate that they do try to personalize shares based on the recipient. To explain these contrasting results, we propose a novel process model of sharing that takes into account people's preferences and the salience of an item.  We also present encouraging results for a sharing prediction model that incorporates both the senders' and the recipients' preferences.   
These results suggest improvements to both algorithms that support sharing in social media and to information diffusion models.
\end{abstract}




\section{Introduction}
Many people get recommendations for movies, music, articles, and products through their social connections both online and off.  Online, we often think of sharing primarily as a public broadcast through tweets, status updates, and the like.  Much online sharing, however, is narrower, targeted at specific audiences (as with Google+ circles or Pinterest boards) or \textit{directed} \cite{bernstein10} at specific individuals through email, chat, and person-to-person messages (e.g. suggesting movies on Netflix \cite{netflix14}). Recent studies show that sharing content through email is still popular \cite{bernstein10,sharethis14}, surpassing social media in certain product categories \cite{socialtwist13}.  

Such sharing is a surprisingly complex process affected by a number of considerations, as illustrated by a study participant:
\begin{quote}
\textit{``I tend to share ... when I understand something about the other person and I think that a certain movie, book, song, etc., might interest that person, whether it be to challenge what someone is saying or feeling, or to reinforce and reaffirm what someone is thinking or feeling. Sure, with close friends, with whom you maintain a close relationship, you might just say, `hey I liked this movie. check it out.' But I think we do that because we already know the person and the person already knows us. There's a certain level of mutual understanding and respect already established. I say I don't recommend `all willy-nilly' and I mean that I don't run up to strangers and recommend they read George Orwell, because I don't know anything about that person or how they feel. Unfortunately, to some extent, we do do just that, we all do that sometimes, when we recommend in order to show off our own interests, to show how cool we are, to show how much we know, to show how diversified our interests are, to show how much niche-specific music we listen to. You know, when we're self-interested assholes.''} (P23)
\end{quote}

Untangling this complexity is important for understanding how information diffuses online.

Recommender systems within social networks will also benefit from better models of sharing decisions. These models can be used to support sharing online by suggesting which items to share and who to share them with \cite{bernstein10,wang13}. 
However, while there is extensive research on understanding people's rating behavior and predicting their preferences for items \cite{susurvey09}, little is known about people's online sharing behavior and its predictability.

From past research on word-of-mouth product sharing and information sharing on the web, we know that people share items for many reasons: enhancement of personal image, personal interest in the item, helping others, a desire to help or harm the item's producer, seeking advice, and so on \cite{sundaram98,dichter66,hennig-thurau03}. On balance, these motivations can be seen as special cases of two primary drivers of sharing proposed by Ho and Dempsey: \textit{individuation}, the need to establish a distinct identity for oneself, and \textit{altruism}, the desire to help others \cite{ho10}.

Based on the primary motivations of individuation and altruism, we can expect people to share content that is some balance of their own and others' interests. What that balance is, and how it comes to be so, however, is an open question and our main focus in this paper. A recent study on Twitter suggests that the balance is tilted toward the self: around 80\% of people primarily share content about their activities and opinions, while only 20\% share informational content more likely to be useful to others  \cite{naaman10}. 
However, this may be because of the broadcast nature of sharing on Twitter where in the absence of a specific audience, sharing becomes an expression of one's thoughts and ideas \cite{zhao09}.  
When people share to specific recipients, they may be more likely to think about usefulness for the recipient (and thus be more altruistic) than when they share to larger groups \cite{berger14,bernstein10}. 

\textbf{Present Work.}
To study the relative effects of individuation and altruism, we conducted an empirical study where pairs of friends were independently shown the same set of movies and asked to rate those movies and/or share them with their friend.
Ratings for movies by the sharer can be considered as a proxy for her preference in movies, which in turn is expected to reflect her self-image (individuation). 
Similarly, we regard sharing movies that align with the recipient's preferences as other-oriented altruistic behavior. 87 pairs of Facebook friends took part in the study, providing rating and sharing data along with answers to open-ended questions about their sharing behavior.

Our results provide several concrete findings about person-to-person sharing. First, individuation is the dominant factor for sharing items. Shared items are rated significantly higher by senders than items that aren't shared. Further, senders' ratings are significantly higher than recipients' ratings for shared items.

Second, participants describe customizing their recommendations based on the recipient, consistent with results from earlier studies \cite{berger14, bernstein10}.  We argue that these two seemingly contrary results---people claiming to personalize but still sharing items that they themselves like---can be best explained by the following decision process: people choose items to share based on their preferences and context, then decide to share or not depending on the recipient.  We formalize this process as the \textit{preference-salience} model of sharing and provide some evidence for it.

Third, we show that we can (noisily) predict which items a person might share.  A model using sharers' and recipients' preferences for movies along with sharers' promiscuity can predict shares by study participants with more than 75\% precision.  We also find that item characteristics such as average rating and popularity play little role in predicting sharing decisions compared to people's own preferences for an item.

\section{Related Work}
Our goal of better understanding of sharing behavior is situated between existing work around word-of-mouth sharing, recommender systems,
and diffusion in social media. We discuss each in turn, along with a key question related to that body of work.

\subsection{Word-of-mouth sharing}
When communicating with others, people customize their message based on their estimation of the audience's knowledge or attitudes \cite{krauss91}. At the same time, much word-of-mouth sharing is driven by people's desire to share items that closely align with or enhance their self-image \cite{taylor12marketing,chung06}.  Our first question addreses how people balance customization with self-expression.

This dual motivation between \textit{self} and \textit{other} has also been found to drive sharing activities in online contexts. Research in online word-of-mouth referrals has shown individuation and altruism as two dominant motivations for sharing \cite{ho10,hennig-thurau03,sundaram98}. Studies of knowledge sharing in online professional communities reveal a similar pattern: people share knowledge to enhance their professional reputation or when they enjoy helping others \cite{wasko05}.
 
When sharing items such as movies, these two motivations of individuation and altruism can be mapped to sharing based on one's own interests or the audience's.  These, in turn, can be estimated from the rich preference data available online, then used to study the relative influence of these factors. 

\noindent \textbf{RQ1: } To what extent do people tend to share items that they like themselves (individuation) versus those that they perceive to be relevant for the recipient (altruism)? 

\subsection{Directed recommendation}
Studying sharing behavior also allows us to ask how well people can recommend content for others; such directed suggestions can provide a useful complement to algorithmically-generated recommendations \cite{bernstein10}.  Most studies in this space have focused
on the question of influence, using recipients' acceptance of recommendations as a proxy for how influential the sender is. For example, network influence \cite{domingos01, kempe03diffusioninfluence}, the relationship with the sender \cite{bond12,bakshy12weakties}, the explanation accompanying the content \cite{sharma13,taylor13,kulkarni13}, and the susceptibility of an individual towards shared items \cite{aral12-susceptible} have all been shown to affect people's likelihood of accepting suggestions.

However, little is known about whether people make suggestions that receivers would actually like.  The study most related to this question compared people's ability to predict a stranger's movie ratings based on part of that person's rating profile to predictions from a standard collaborative filtering algorithm \cite{krishnan08}. On balance, people were not as accurate---and, interestingly, got worse as the profile became more similar to their own.

In this paper, we study how well people's suggestions match the recipient's interests when they share items with known friends and compare their results with algorithmic recommendations.  Instead of making inferences from profile information \cite{krishnan08}, people rely on their own knowledge about a friend to choose which items to share, which we see as a more natural recommendation scenario.  For consistency in terminology, we use recommendations to refer to algorithmic recommendations and shares to refer to human-generated directed recommendations in the rest of the paper.

\noindent \textbf{RQ2:} Do people share items to friends that are well-liked compared to recommendations?

\subsection{Diffusion models} In computer science, sharing is most commonly studied as a component of information diffusion models \cite{watts02simplecascades,kempe03diffusioninfluence, bakshy12weakties,gruhl04,cha09}. These models simulate the spread of items in a network, where people adopt an item through either probabilistic transfer between connected people or based on a threshold number of adoptions in a person's neighborhood. However, these models don't actually explain most adoption in social media \cite{goel12} because the viral analogy breaks down.
Sharing is a \textit{voluntary} process shaped by social forces such as 
people's willingness to diffuse \cite{lagnier13probdiffusion}, attention to targets' needs \cite{bernstein10}, and relations between sharer and target including tie strength \cite{brown87} and homophily.

The present work aims to build models of sharing that account for this voluntary decision-making. % and uses characteristics of the sharer, the receiver, and items. 
If successful, these models can explain how people weigh their own and recipients' preferences when sharing items and present a computational framework for predicting future shares. This model can be used to estimate sharing probabilities for different recipients and items; diffusion models can leverage these probabilities to better account for these influences and make more accurate predictions around diffusion. 

\noindent \textbf{RQ3: } How well can we predict whether an item is shared using readily available information about people's preferences and properties of people and items?

\begin{figure}[tbh]
\center
\includegraphics[scale=0.64]{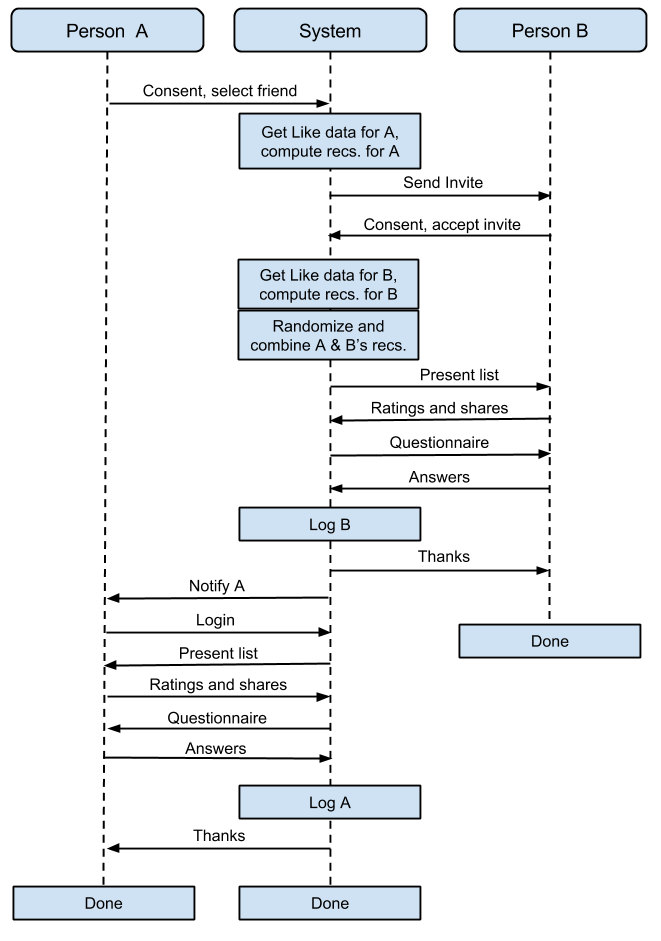}
\caption{The flow of the experiment. Person A invites a Facebook friend (B) to take part in the study. Once B accepts, B rates and shares recommendations computed from both A and B's past movie Likes. Finally, person A logs into the study again and rates and shares an identical set of recommendations.  To reduce effects of social influence, there is no direct communication between A and B and the system presents no information about the other person's decisions.
}
\label{fig-user-flow}
\end{figure}

\begin{figure}[tbh]
\includegraphics[scale=0.55]{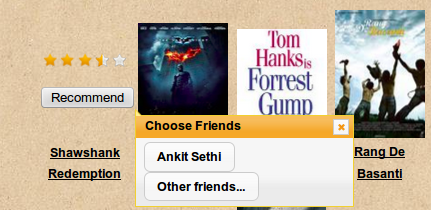}
\caption{A screenshot of the interface. A maximum of 20 movies was shown; participants rate and/or share as many movies as they wish.  The study partner was shown as the default recipient.
}
\label{fig-rec-suggest}
\end{figure}

\section{Description of the user study}
\label{sec-studydesign}
To tackle the above questions, we conducted an experiment that asked pairs of friends on Facebook to rate and share items.  
We chose movies as the item domain for several reasons.  Movies are a common domain in recommender systems research and an important cultural item that people often share and discuss, making them a natural domain for studying sharing.  They are also fairly popular to Like on Facebook \cite{sharma11} (among our participants, $\mu=18.2,\sigma=31.8$), allowing us build reasonable user profiles for  making recommendations.  

\noindent \textbf{Study logistics.} Figure~\ref{fig-user-flow} shows an overview of the experiment, which proceeded in three main stages. In the first stage, a participant (A) signs up for the study and invites one of her Facebook friends as a partner (B) through email sent by our system.  When Person B accepts the invitation, the second stage starts. Person B is shown a set of movie recommendations which he is asked to rate and/or share with Person A, followed by a questionnaire that asks about B's relationship with A and his practices around sharing items in general. Person A then gets a notification and, in the third stage, performs the same tasks on the same items, then answers the same questions as Person B.

Showing both partners the same items allows us to get overlapping sharing and rating data to address the research question.  The asynchronous design allows partners to participate independently, making the study easier to complete.  To minimize explicit social influence that might affect sharing and rating behaviors \cite{sharma13}, participants do not see information about their partner's ratings or shares.  

Figure~\ref{fig-rec-suggest} shows a screenshot of the interface, designed to broadly resemble other systems that recommend lists of items. Participants were free to choose the movies to rate or share among the movies shown.  All ratings are on a Likert scale from
0.5-5, with half ratings allowed.  Movies are shared by clicking on the Recommend button and providing a short message explaining the recommendation.  The system showed the study partner as the default choice for sharing; all but four shares were to their partner so we removed those four shares from the dataset.

After the rating and sharing task, participants completed a short questionnaire asking how close they were to their study partner, as well as open-ended questions about how and why they suggest items, and how and when they receive suggestions from others.

\noindent \textbf{Computing recommendations.} For each user, we computed recommendations based on Likes within their ego network.  We first selected the user's $k=20$ most similar friends based on Jaccard similarity of their Likes with the user. We then computed a score for each movie based on its similarity-weighted popularity among the $k$ friends: 
$$Score(item_i, u) = \frac{\sum_{j=1}^k JSim(u,f_j) Likes(f_j, item_i)}{\sum_{j=1}^k JSim(u,f_j)}$$
where $Likes$ is 1 if friend $f_j$ likes $item_i$ and 0 otherwise.  This algorithm gives comparable results to using data from thousands of Facebook users using a generic k-nearest neighbors algorithm \cite{sharma-icwsm13}.

The ten highest scoring movies for each user that were not already liked by her were chosen as recommendations.  Some users may have less than 10 recommendations because they do not have enough friends or enough Likes in their profile to compute recommendations. Further, 
Facebook API errors and rate limits prevented some users' Likes from being fetched. Thus, participants saw between 0 and 20 movies; we pruned those who saw less than 10. 

Recommendations for both partners were computed and stored in the second stage, ensuring that both saw the same set.  Each pair's recommendations were combined and presented in a randomized order to minimize presentation order effects.

\begin{table}
\center
\begin{tabular}{lrrr}
\toprule[0.10em]
\textbf{Participants' statistics} & \textbf{MTurk} & \textbf{Univ} & \textbf{All}  \\
\midrule
Rated at least once & 36 & 82 & 118 \\
Shared at least once & 28 & 58 & 86\\
Total ratings & 246 &  720 & 966\\
Total shares & 97 & 217  & 314\\
Num. ratings/person  & 6.83 & 8.81 & 8.18 \\
Mean rating/person & 3.82 & 3.87 & 3.85 \\
Num.  shares/person & 2.69 & 2.64  & 2.66\\
Num. likes/person & 16.2 & 19.1  & 18.2\\
%\bottomrule[0.10em]
\end{tabular}
\caption{Aggregate statistics for participants recruited from Amazon Mechanical Turk and the university. About a third of the participants were recruited through Mechanical Turk. There was no significant difference in study activity between the two populations.}
\label{tab-pop}
\end{table}

\noindent \textbf{Participation.} We recruited participants through two sources, a pool of participants at a large northeastern U.S. university and Amazon Mechanical Turk. The university pool consists of students and staff who elect to take part in user studies.  We conducted a drawing with a 1/3 chance of winning \$10 gift cards to motivate participation inside the university and paid Mechanical Turk users a flat \$2.50. There were no significant differences in terms of the number of shares, ratings, or Facebook Likes between the groups so we treat them as a composite sample (Table~\ref{tab-pop}).

\begin{figure}
	\centering
	\includegraphics[scale=0.5]{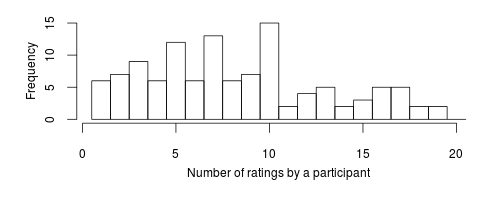}
	\vspace{-8 mm}
	\includegraphics[scale=0.5]{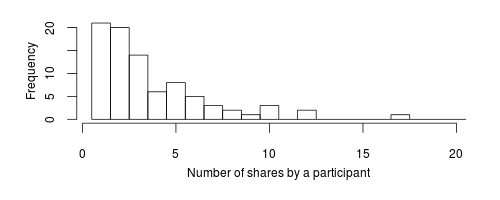}
	\vspace{2 mm}
	\caption{Distribution of ratings and shares per participant.  On average, people rated about three times more items than they shared.}
	\label{fig-rate-share-distr}
\end{figure}

After pruning people who saw fewer than ten recommendations, a total of 87 pairs took part, 59\% female. Due to turnover between the three stages, only 142 participants saw recommendations. Figure~\ref{fig-rate-share-distr} shows the distribution of number of ratings and shares by the participants.  118 participants rated at least one movie and 86 shared at least one movie for a total of 966 ratings and 314 shares to their partner; each session took 11 minutes on average. These are the data that we consider for our analysis.

We expected pairs to know each other (and their preferences) well since people chose their own partners. When asked to evaluate the statement ``We are very close to each other", 83\% of participants answered ``Agree" or ``Completely Agree", indicating that most pairs were close ties.

Participants were divided into three groups based on the recommendations they saw during the experiment.
\begin{itemize}
 \item \bothshown: All participants who saw more than 10 movies belong to this group. Since our algorithm computes a maximum of 10 recommendations for each user, this means they saw movies recommended based on both their own profile and their partner's.  The least number of movies shown is 14 for this group.
 
 \item \ownshown: These participants saw 10 movies that were recommended based on their own profile. 
 
  \item \othershown: These participants saw 10 movies that were recommended based on their partner's profile. Since both participants in a pair see the same movies, this means that partners of participants in \ownshown\ are in \othershown\ and vice versa.
\end{itemize} 

Table~\ref{tab-conditions} shows the breakdown of participants between the three groups. Note that these groups are not randomly assigned; Liking behavior and API errors in data fetching affected the recommendations any participant saw and thus which group they are in. 

Before we discuss factors affecting sharing, a couple of sanity checks for our study design are in order. The first concerns the efficacy of our recommendation algorithm. The average rating of items recommended using a participant's own Likes is significantly higher than those recommended for her partner ($\mu=3.93, \sigma=1.00, N=515$; $\mu=3.76, \sigma=1.17, N=451$; $t(887)=2.37, p=0.02; d=0.2$\footnote{We report Cohen's-d \cite{cohen92} as a measure of effect size throughout.}). This indicates that the algorithm does capture users' preferences to some extent. 

Second, while we designed the study so that participants do not have an incentive to tell their partners about their shares (compensation was for completing the study, not getting the ratings right), nothing prevents participant B from disclosing her shares to A before A logs on to the study again (especially if they are close to each other). To check whether our study results may have been impacted by such information exchange, we compared the average ratings for received movies (i.e., those that were shared by their partner) between participants who completed the experiment first and those that completed the experiment after their partner. A t-test revealed no significant difference ($\mu_A=3.87$, $\mu_B=3.89$), which makes us believe that such disclosure between A and B was not prevalent.

\begin{table}
\center
\begin{tabular}{lccc}
\toprule[0.10em]
\textbf{Condition} & \textbf{No. of Users} & \textbf{Ratings} & \textbf{Shares} \\
\midrule
\bothshown & 60  & 609 & 141 \\
\ownshown & 29 & 179 & 96 \\
\othershown & 29 & 178 & 77\\
\midrule
\allcond & 118 & 966  & 314 \\
\end{tabular}
\caption{Aggregate rating and sharing statistics for users in the three groups.
\bothshown\ participants saw a mix of recommendations for themselves and their partner, \ownshown\ participants saw only recommendations made for themselves, and \othershown\ only saw recommendations made for their partners.}

\label{tab-conditions}
\end{table}

\begin{figure}[tb]
	\center
	\includegraphics[scale=0.5]{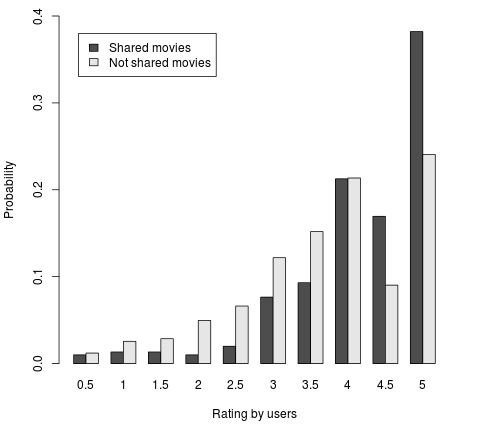}
	\caption{Probability of different sharer ratings for shared and non-shared items. Highly rated items are more likely to be shared than lower-rated items.}
	\label{fig-shared_notshared}
\end{figure}

\section{The role of own preferences in sharing}
We start with \textbf{RQ1}, examining the extent to which people share items they like themselves versus those they perceive to be relevant for the recipient. We use both people's sharing data and their open-ended answers to the question: ``How/why do you suggest items to people?'' 

\begin{table*}
\center
\begin{tabular}{lrccrcclc}
&& Shared  &&& Non-Shared && & \\ 
%\toprule[0.15em]
\cmidrule(r){2-4}
\cmidrule(r){5-7}
\tabhead{Group} & $N$ & $\mu$ & $\sigma$ & $N$ & $\mu$ & $\sigma$ & Significance & Effect Size \\ 
\toprule[0.10em]
\bothshown & $140$ & $4.20$ & $0.93$ & $469$ & $3.81$ & $1.07$ & $t(258)=4.20; p<0.0001$ & $0.4$\\
\ownshown & $90$ & $4.16$ & $0.87$ & $89$ & $3.56$ & $1.05$ & $t(170)=4.14; p<0.0001$ & $0.6$ \\
\othershown & $71$ & $4.18$ & $1.08$ & $107$ & $3.31$ & $1.24$ & $t(164)=4.92; p<0.0001$ & $0.7$ \\
\midrule
\allcond & $301$ & $4.18$ & $0.95$ & $665$ & $3.70$ & $1.11$ & $t(671)=6.98; p<0.0001$ & $0.5$ \\
\end{tabular}
\caption{Comparison of sender ratings for shared and non-shared movies, along with results of an unpaired $t$-test and Cohen's-d effect size measure. In all three groups, shared movies are significantly higher rated than non-shared movies.}
\label{tab-ownpref_sig}
\end{table*}

\subsection{Senders' own preference matters}
We first analyze whether senders tend to share movies they like by comparing their ratings for shared and non-shared movies.  The overall distribution of ratings for shared and non-shared movies is shown in Figure~\ref{fig-shared_notshared}. On average, shared movies are rated higher.  
Further, highly rated movies are more likely to be shared: 77\% of the shared movies are rated 4 or above.

Table~\ref{tab-ownpref_sig} shows that for all three groups of participants, shared movies are rated significantly higher than those that are not shared\footnote{Ratings are not strictly normal and independent, which are assumptions for conducting a t-test. There is a skew towards higher ratings and ratings for the same movie or by the same user may be interdependent. Thus, we also considered a linear mixed-effects model to account for sender and item variability as a random effect with sharing as a fixed effect. The results were similar and are described in the Appendix.}. The effect size is biggest for the \othershown\ group.  This is because the mean rating for shared movies ($\mu=4.18$) is as high as for other groups, but the mean rating for non-shared movies is lower since the recommendations were generated for the sharer's partner rather than the sharer.

\subsection{Item characteristics are not informative}
One possible explanation for shared movies having higher ratings is that sharers may only share high quality or popular movies. To check this, we collected average rating and popularity data from the popular movie reviewing website IMDB\footnote{Internet Movie Database  www.imdb.com, accessed Feb. 2014.}. For each movie, IMDB reports an average rating on a scale of 1-10 (\textit{IMDB rating}) and the number of people who have rated that movie (\textit{IMDB popularity}). 

We find that there is no significant difference between shared and non-shared movies for either IMDB rating ($\mu=7.64, \sigma=1.23; \mu=7.48, \sigma=1.20$) or IMDB popularity ($\mu=11.2M, \sigma=9.68M; \mu=13.1M, \sigma=10M$).
This indicates that aggregate opinions about items such as average rating or popularity did not matter much when sharing movies. These results align well with the motivation of individuation, which suggests that people will share items that help establish a distinct identity for themselves.

\subsection{Participants' responses support individuation}
These results are supported by participants' accounts of how they select items to share.
For more than half of the participants, liking the item themselves is the most important factor in sharing an item with someone. 
\begin{quote}
\textit{``Usually when I suggest, it depends on the item, not the target individual, because I want to share what I enjoyed.''} (P8)
\end{quote}
Sharing items that one likes may also signal expertise \cite{wasko05}.
\begin{quote}
\textit{``I suggest items to people because in my case, I've usually seen more movies than they have and I have a better relative perspective of what is considered good or bad.''} (P61)
\end{quote}
It can also be a useful way to have shared experiences and discussions around items.
\begin{quote}
\textit{``I suggest because I like something and I want to see if other people feel the same way about an item. When I suggest items to my friends we are able to talk and laugh about the certain item.''} (P91)
\end{quote}
All of the above can be connected to individuation as the guiding motivation for sharing.

\subsection{Differences in sharing promiscuity among participants}
While participants tended to share movies that matched their preferences, we saw great variability in how much people shared, or their sharing \textit{promiscuity}. Among the users who shared movies, the minimum number of movies shared was 1 and the maximum was 17 ($\mu=3.65, \sigma=3.05$), as shown in Figure~\ref{fig-rate-share-distr}.

For many people, sharing is reserved only for \textit{``something I really, really enjoy''} in part because sharing too frequently \textit{``tends to water down my stamp of approval.''} (P16)

Selective sharing is also connected to the common problem of managing one's image in social media \cite{taylor12marketing}.
\begin{quote}
\textit{``Sometimes I'm paranoid that if I suggest items to someone I don't know very well, they will change their perception of me.''} (P15)
\end{quote}

A likely hypothesis connected to sharing promiscuity is that people tend to select items for sharing in decreasing preference order.  This suggests that sharing more items should lead to lower average ratings by both senders and recipients; in fact, there is a negative correlation between the number of items shared and both average sender ($corr=-~0.31$) and recipient ($corr=-~0.36$) ratings.

\begin{figure}[tb]
	\center
	\includegraphics[scale=0.5]{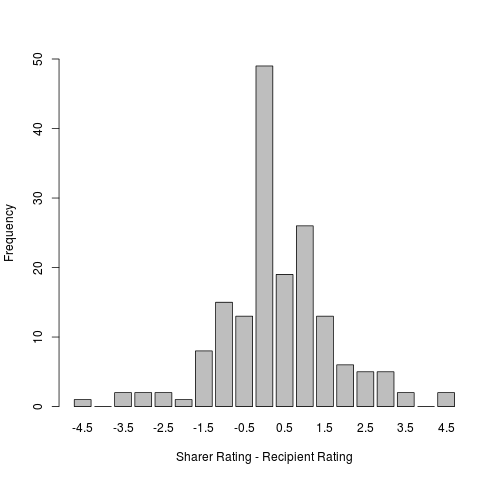}
	\caption{Comparison of sender and recipient ratings for shared movies. The x-axis represents the difference between the sender and recipient rating; on average, sender ratings are higher than recipient ratings.}
	\label{fig-recipient-sender-hist}
\end{figure}

\begin{table*}[tb]
\center
\begin{tabular}{lrccrcclc}
&& Sender Rating  &&& Recipient Rating && & \\ 
\cmidrule(r){2-4}
\cmidrule(r){5-7}
\tabhead{Group} & $N$ & $\mu$ & $\sigma$ & $N$ & $\mu$ & $\sigma$ & Significance & Effect Size \\
\toprule[0.10em]  
\bothshown & $81$ & $4.12$ & $0.95$ & $81$ & $3.80$ & $1.05$ & $\ t(80)=2.19; p=0.01$ & $0.3$ \\
--\ownalgo & $38$ & $4.14$ & $0.94$ & $38$ & $3.71$ & $1.27$ & $\ t(37)=1.93; p=0.03$ & $0.4$ \\
--\otheralgo  & $43$ & $4.10$ & $0.97$ & $43$ & $3.88$ & $0.81$ & $\ t(42)=1.14; p=0.13$ & $0.2$ \\
\ownshown & $49$ & $4.40$ & $0.75$ & $49$ & $3.67$ & $1.34$ & $\ t(48)=3.52;p<0.001$ & $0.7$ \\
\othershown & $41$ & $4.06$ & $1.09$ & $41$ & $4.28$ & $0.70$ & $\ t(40)=1.10; p=0.14$ & $0.2$ \\
\midrule
\allcond & $171$ & $4.19$ & $0.94$ & $171$ & $3.88$ & $1.09$ & $t(170)=2.90; p=0.002$ & $0.3$ 
\end{tabular}
\caption{Comparison of sender and receiver ratings for shared movies using a paired t-test. Across all groups, shared movies have a significantly higher rating from the sender than the recipient when the sender shares from a list close to her movie preferences. The difference is not significant when a sender shares from a list close to the recipient's movie preferences; still, sender ratings in this case are high ($\mu>4$). }
\label{tab-sender_receiver_sig}
\end{table*}

\section{How useful are shares for the recipient?}
Analysis of recipients' ratings for shared items reveals more about the relative effects of individuation and altruism (\textbf{RQ1}).

\subsection{Senders rate shared items higher than recipients} 
A total of 171 shares were rated by both sharers and recipients.
Figure~\ref{fig-recipient-sender-hist} shows the difference between their ratings.  About half of the shares have a higher sender rating and a quarter have equal ratings. A paired t-test for these shares shows that senders' ratings are significantly higher than recipients' ($\mu=4.19,\mu=3.88; t(170) = 2.90; p=0.002$).

Although recipients' ratings for shared items are lower than senders', %and no better on average than for algorithmic recommendations, 
many participants claimed that they consider the recipient's preferences before sharing an item. 
\begin{quote}
\textit{``I make suggestions to people if I think they might gain enjoyment. Obviously it really depends on their personality and their likes/dislikes.''} (P22)
\end{quote}
This disconnect between people's self-reports and actual sharing behavior is surprising; we will consider likely explanations for it in the discussion.

\subsection{Recipients' ratings depend on the item set shown}
When senders saw recommendations from both algorithms (\bothshown\ group), sender rating is significantly higher than the recipient rating for a shared item (Table~\ref{tab-sender_receiver_sig}). 
However, when we break up the shares in the \bothshown\ condition by algorithm, a more complex picture emerges. Although participants shared movies about equally from both sets of recommendations, the difference in sender and receiver ratings is significant only for the movies selected by sender's \ownalgo.

We see a similar pattern when we compare the \ownshown\ and \othershown\ groups. As shown in Table~\ref{tab-sender_receiver_sig}, participants in the \ownshown\ group had significantly higher ratings than the recipient, but not those in the \othershown\ group. In fact, recipients' ratings were higher than senders' ratings for shares in the \othershown\ group.

It is not surprising that recipient ratings are higher when shares come from movies recommended by \otheralgo\ because those recommendations are based on the recipient's past Likes. Still, senders' rating for shares is high across groups and algorithms. These findings, coupled with higher ratings by senders for shares versus non-shares, lead us to conclude that one's own preferences (and thus individuation) are the dominant criterion when choosing movies to share.

\begin{table*}[tb]
\center
\begin{tabular}{lrccrcclc}
&& Shared  &&& Recommended && & \\ 
\cmidrule(r){2-4}
\cmidrule(r){5-7}
\tabhead{Group} & $N$ & $\mu$ & $\sigma$ & $N$ & $\mu$ & $\sigma$ & Significance & Effect Size \\
\toprule[0.10em]  
\bothshown & $81$ & $3.80$ & $1.05$ & $119$ & $3.81$ & $0.97$ & $t(164)=0.05; p=0.95$ & $0.04$ \\
\ownshown & $42$ & $4.28$ & $0.70$ & $109$ & $3.86$ & $1.01$ & $t(103)=2.75; p<0.01$ & $0.5$ \\
\midrule
\textit{Combined} & $123$ & $3.96$ & $0.96$ & $228$ & $3.84$ & $0.97$ & $t(250)=1.07; p=0.28$ & $0.1$ \\
\toprule[0.10em]  
\end{tabular}
\caption{Comparison of recipients' ratings for shared and recommended movies. Shared movies are rated significantly higher when senders were shown only movies tailored for the recipient, not when they were shown mixed recommendations. Recipients in the \othershown\ group did not see recommendations tailored for them, so we excluded them from the comparison.}
\label{tab-share_algo_compare}
\end{table*}

\subsection{Shares are comparable to algorithmic recommendation}
We saw that when restricted to the movies recommended for recipients, people share movies that are more aligned with recipient's preferences.  We now turn to \textbf{RQ2}, about how those shares compare to the recommendations generated by the algorithm for the study.

We use data from the two groups (\bothshown\ and \ownshown) where  recipients saw recommendations based on their own Likes and compare these recommendations with shares. For the \bothshown\ group, about half of the movies shown were recommended based on the recipient's preferences.  A t-test shows that the recipients' average rating of shared movies is not significantly different than their average rating for algorithmically recommended movies (see Table~\ref{tab-share_algo_compare}).
%Among these, we compare the ratings given by recipients for the movies that got shared to them versus all of them.
This indicates that shares are comparable to a recommender algorithm in terms of recipient liking. 

For recipients in the \ownshown\ group, whose partner shared movies from a set of recommendations tuned for the recipient, the average rating for shared movies is significantly higher than for algorithmically recommended ones.
Thus, items deemed relevant by the recommendation model attain  higher ratings if they are also shared by a person's friend. 

\subsection{Shares have higher overall quality than recommendations}
However, shared items do differ from recommended items in terms of overall quality.  We again use IMDB rating and IMDB popularity as measures of movie quality and popularity.  Among all movies (with or without participant ratings), shared movies are of higher average IMDB rating than recommended movies ($\mu=7.83,\sigma=1.03$; $\mu=7.54,\sigma=1.15$; $t(317)=3.02, p=0.0026$; $d=0.3$), but there is no significant difference in popularity ($median=9.65M$, $median=9.52M$). This indicates that on average people select higher quality movies to share compared to the baseline recommender.

\section{Predicting shares}
From the last two sections, it seems that senders' own preferences for movies matter more than the recipients' in sharing decisions, although they are still at least as well-liked by recipients as recommendation.
We now examine how well we can predict these sharing decisions (\textbf{RQ3}) using information about senders, recipients, and items.
We build a series of models---starting from simple ones that use a single feature---to predict shares. 

\noindent \textbf{Data and method.}
We have 279 shares for movies that also have IMDB rating and popularity data.  We use these shares to create 10 balanced datasets by randomly sampling sets of 279 non-shares.
For each model, we perform 10 cross-fold validation in each dataset and average the results.  For ease of interpretability, we use a decision tree classifier\footnote{We also tried random forests, logistic regression, and support vector machines.  Results were qualitatively similar.}. 

\noindent \textbf{Computing features.} Based on our results earlier, we consider a sharer's own preferences for an item, her sharing promiscuity, and the recipient's preferences for the item as features for prediction.  We also add preference similarity between the sharer and recipient to examine effects of homophily, along with IMDB rating and IMDB popularity to examine effects of item characteristics.

Sharing promiscuity is the number of shares by a user in our training dataset. 

The sharer's and recipient's rating are not available for every item, so we estimate their preference for a movie through a method similar to item-based collaborative filtering \cite{sarwar01}.
We convert ratings in the study to a unary scale by denoting each rating 4 or above as a Like and combine those with the Likes we collected during the study, giving a total of 43K users and 785K likes on all movies.
We represent each movie as a set of users who Liked the movie and compute Jaccard similarity between each pair of movies.

To estimate a user's preference for a movie, we compute the average similarity between the given movie and the movies that a user had Liked.  We use this similarity score between a user and a movie as a feature denoting their preference for the movie, computing both the sender's and recipient's preference for each movie.

Finally, we compute the sender-recipient similarity feature as the Jaccard similarity between sets representing each user's movie Likes as defined above. % and unary ratings by each user.

\begin{table}[tb]
\center
\begin{tabular}{l@{\hskip -0.03in}ccc@{\hskip -0.01in}}
\toprule[0.10em]
\textbf{Features}  & \textbf{Precision} & \textbf{Recall} & \textbf{Accuracy} \\
\midrule
\textbf{Item-based} & & &\\
\ Average IMDB Rating  & 49.5 & 61.9& 50.1 \\
\ Popularity &  51.8  & 60.6& 51.1\\
\ Both  & 51.8 &62.1& 50.9 \\
\textbf{Recipient-based} & & & \\
\ Recipient-Item Similarity  & 64.0 & 38.5& 58.7\\
\ Sender-Recipient Similarity  & 64.8 &41.2 & 58.7\\
\ Both  & 62.9 & 54.1& 60.5 \\
\textbf{Sender-based} & & &\\
\ Sender-Item Similarity  & 66.3 & \textbf{79.2}& 68.4\\
\ Sharing Promiscuity & 69.0  & 72.1& 69.1 \\
\ Both  & 72.3 & 74.9& 72.7 \\
\midrule
\textbf{Sender+Recipient}  & \textbf{78.4} & 70.8&\textbf{75.7} \\
\end{tabular}
\caption{Precision, recall, and accuracy for predicting whether an item is shared. Bold numbers are per-metric maximums. Item features such as popularity and average rating do little better than random guesses.  Recipient-based features improve precision, but the most predictive features are connected to the sharer.}
\label{tab-predict-shares}
\end{table}

\noindent \textbf{Prediction performance.}
Table~\ref{tab-predict-shares} shows each model's precision, recall, and accuracy, common metrics for evaluating such models. Accuracy is the overall fraction of correct predictions of whether an item is shared or not. 
Sometimes it makes sense to focus only on predicted or actual shares; to do this, we also compute precision and recall.  
Precision is the fraction of correct predictions among all the items predicted as shares and recall is the fraction of true shares that were correctly predicted.

Item-based features of movies such as quality or popularity have little predictive power, with accuracy close to the 50\% that a random predictor would achieve on a balanced dataset.  

Using only recipients' similarity with a movie gives a precision of 64\%, but the recall is low (38\%). This is because a high recipient rating is a better than random predictor of a share but does not cover many other shares that have lower recipient ratings.  The similarity between Likes of the sharer and the recipient gives comparable precision and recall to using recipients' similarity. 
 
Sender-based features are more useful. A sender's similarity with a movie is able to predict whether a movie is shared or not with 66\% precision and 79\% recall, higher than recipient-based features. These results are consistent with the results around individuation described earlier. 

Sharing promiscuity of a sender is also important; shares can be predicted with 69\% precision based only on promiscuity. The model, though, is trivial, predicting that users above a certain threshold of promiscuity would share all movies shown to them while those below the threshold will share none.

The model that includes both promiscuity and similarity (the sender-both line in the table) is more interesting.  Precision increases to 72\% compared to either alone; a fitted decision tree is shown below. Similar movies above a threshold are shared depending on the sharer's promiscuity, but not those below it.

\begin{verbatim}
sharer_sim <= 0.0101: Non-shared
sharer_sim > 0.0101
| sharer_prom <= 1: Non-shared
| sharer_prom > 1: Shared
\end{verbatim}
   
Finally, combining sender-based and recipient-based features achieves an accuracy of 76\% and precision of 78\%.  Knowledge about preference similarity between the sender and the recipient helps; however, the decision tree ignores a recipient's similarity to the item.
\begin{verbatim}
sharer_sim <= 0.0107: Non-shared 
sharer_sim > 0.0107 
| sharer_prom <= 3
| | sharer_prom <=1: Non-shared
| | sharer_prom > 1
| | | sharer_recip_sim <= 0.0754
| | | | sharer_sim <= 0.0601: Non-shared
| | | | sharer_sim > 0.0601: Shared
| | | sharer_recip_sim > 0.0754: Shared
| sharer_prom > 3: Shared
\end{verbatim}

Although our experiment design restricted the set of movies that can be shared and we used a modified, balanced dataset of shares and non-shares, these results demonstrate the potential of predicting sharing decisions using people's preferences. 

\section{Discussion}
When broadcasting as on Twitter, past research shows that most people post messages about themselves rather than sharing useful information \cite{naaman10}. Based on results in communication around tuning messages for the audience \cite{krauss91} and recent work showing that people think more about usefulness for the recipient as the audience size decreases \cite{berger14}, we expected people would weigh recipients' preferences more when sharing to an individual.  However, both people's sharing data and their self-reports underscore the importance of their own preferences. For \textbf{RQ1}, the answer is clearly that sharing is more driven by people's own preferences than recipients'.   

Yet people claimed to customize their shares for the recipient.
Comparing shares with recommendations (\textbf{RQ2}) suggests that the items shown to a person affected their sharing behavior.  When restricted to a set of items recommended for the recipient, people share items that are on average better liked than the recommendations, but not when they are shown their own recommendations or a mix of both. 

Further, much of the sharing in the experiment could be predicted using people's preferences (\textbf{RQ3}). In particular, sharer-based features such as similarity with the item and promiscuity are important and can predict shares with over 70\% precision.  Recipient-based features are useful but not as discriminating, while item-based features reveal little.

To explain these findings, we propose a model for sharing and discuss the implications of that model for improving both information diffusion models and practices of online sharing.

\subsection{A preference-salience sharing process} One way of explaining the disconnect between people's data and descriptions of how they personalize shares for recipients 
is that people do not really try to balance individuation and altruism when they share items.  Rather, they share based on their preference for items and what is salient to them at the moment.  Here salience denotes the particular items and recipient that the sharer is thinking of. 

\begin{quote}
\textit{``I try to assess if the individual that I am recommending to would like the movie that I am suggesting. Otherwise, I do not tell them about the movie, and may think of someone else who would like the movie.''} (P5)
\end{quote}

In addition to their own preferences, people's selection of a candidate item for sharing also depends on the context that makes a certain item salient. When asked ``When do you suggest items to others?", participants responded that they share just after consuming an item, during conversations when a relevant topic comes up, or when asked explicitly.
\begin{quote}
\textit{``I usually suggest either after I see the content or if something related comes up in conversation.''} (P82)
\end{quote}

Thus, a likely process for sharing can be described as follows. People's own preferences determine shareable items.  Among these candidates, some items become salient based on the context and then are shared or not depending on whether the sharer thinks they are suitable for the recipient\footnote{This is somewhat the dual of the FeedMe system's making possible recipients for an item salient by recommending them as targets \cite{bernstein10}.}.

This process can explain how participants shared items that they like, yet claim to be personalizing for the recipient. Out of the movies shown, participants considered the movies that they like for sharing, and then decided to share or not in part based on their perception of their partner's preferences.  The increase in recommendation quality for shares when selecting from items tuned to recipients underscores the saliency aspect: showing items appropriate for the sharing task led to shares that recipients rated higher. 

We believe our preference-salience model presents a reasonable abstraction of people's sharing decisions that has more empirical support than two other models we considered:
\begin{itemize}
\item \textit{High Quality Model. }It is possible that people simply share higher quality items which are likely to be liked by all. This is supported by the fact that shares are rated highly by the senders, are comparable to recommended items in matching recipients' preferences, and are significantly higher-rated on IMDB than recommendations. 
However, there is no difference between overall IMDB ratings for shared and non-shared movies, which led us to reject this model.

\item \textit{Misguided Altruism Model.}  It could also be that people do try to customize shares to recipients but fail because of imperfect knowledge \cite{krauss91}. This is supported by participants' accounts of how they personalize for recipients and the fact that shared items are not rated as highly by recipients as they are by sharers.
However, across all groups of participants, senders' own ratings are significantly higher for shares than for non-shares, which indicates that even if people do try to personalize for the recipient, their own preferences still play an important role.
\end{itemize}

\subsubsection{Still, a simplification of a complex process}
While we propose the preference-salience model as a likely explanation of our observations, our binarization of motivations into individuation and altruism is a simplification that  does not account for other motivations (e.g., to dissuade people from trying an item) or factors (e.g., relationships between people) that affect sharing behavior.

In particular, the closeness of ties between most of our participants might have played a role in people's decisions. 
Knowing a recipient well increases people's chances of knowing his preferences and customizing their suggestions.
\begin{quote}
\textit{``I'll only recommend a movie to a close friend or relative because I know them well enough to know what they would like in a movie.''} (P54)
\end{quote}
Close ties may also allow people to be more open about their preferences.
\begin{quote}
\textit{``With my close friend I feel like I can share anything, but with an acquaintance, I will feel less open to sharing my interests.''} (P71)
\end{quote}

Finally, people don't just share good items; sharing may also warn others about bad items.
\begin{quote}
\textit{``if i really liked something i want others to experience it too...if i hated it i want to help them avoid it.''} (P34)
\end{quote}

\subsection{Applications of the preference-salience sharing process} 
We now discuss how the preference-salience process of sharing can be used to improve diffusion models and online sharing practices.

\paragraph*{Directed sharing in diffusion models}
By demonstrating the effect of people's preferences on directed sharing, our work joins other recent work in questioning the broadcast assumption used for modeling diffusion in social networks \cite{goel12,socialtwist13}. 
One way to improve these models is to acknowledge that not all sharing is broadcast; as described in the introduction, much sharing takes place to individuals and small audiences (@-mentions in Twitter, Google+ circles, small subsets of friends in Facebook).

The preference-salience model lends itself well to incorporating sharing decisions in a diffusion model. Instead of assuming that items are shared to each friend with a fixed probability, the model posits that salient items that match senders' preferences are more likely to be shared, and (to a lesser extent) only to recipients who are more likely to like it.  Salience, preferences, and similarity could be measured and used to set sharing probabilities tuned for each sharer, recipient, and item, which we expect would lead to more accurate diffusion models.

Accounting for sharing promiscuity of individuals is also important, as demonstrated in a recent study on diffusion within blog networks \cite{lagnier13probdiffusion}. Although we did not consider promiscuity in our preference-salience process model, it can be readily included. 
Our process model essentially defines which items are good candidates for sharing. Promiscuity can be thought of as a cutoff for sharing: people with low promiscuity would have a higher barrier for actually sharing one of the candidate items.  We can estimate promiscuity by considering the number of shares made by a user, either absolute or relative to others, as we did for our sharing prediction model.

\paragraph*{Supporting effective sharing}
Our work also has implications for improving sharing behaviors in social networks.  We found that higher promiscuity is correlated with lesser liked shares by both senders and recipients. Systems might help people consider this by providing feedback about their promiscuity relative to others.  Similarly, showing information about the outcomes of their shares such as the average liking by recipients might help people reflect on how their shares are being received. 

The model suggests that making the right items salient may also improve sharing outcomes. 
In personal consumption contexts, systems should still focus on recommending items that align with the user's interests.  In contexts where people are likely to share items with friends or consume them together, however, our results suggest that showing recommendations tailored to the friends rather than the user might lead to better outcomes.

\subsection{Limitations}
It is important to keep in mind the limitations of our study while interpreting the above results.  First, although we broadened our sample by recruiting from two different participant pools in the U.S., differences in demographics and culture may affect people's decisions around sharing. 

The design of our experiment may also have affected people's sharing behavior. We allow participants to invite their own partners so they can choose people with whom they feel comfortable sharing and whose preferences they would be more likely to know.  This led to high tie strength for most of the participant pairs; understanding more about sharing between weaker ties would be an interesting area to study.

We also restricted the set of items in order to get ratings from both members of a pair.  Many real contexts also make subsets of an item domain salient, such as in recommendation lists and filtered activity feeds, and we expect our results apply best there.  Studying scenarios where items are not made explicitly salient (e.g., searching for an item and sharing) would tell us more the relative effects of salience and personal preferences.

Finally, we studied movies as a specific domain.  Though a reasonable choice, sharing decisions in other domains such as news or photos might be different because of differences in cost of consuming an item or ease of sharing items.  Our results may also less readily apply to knowledge sharing scenarios where goals may be more strategic and individual preference may be expected to be less discriminating (such as sharing job information or advice). 

\section{Conclusion}
Our results demonstrate connections between people's preferences and their sharing behavior that pave the way for better models of sharing in online social networks.  Individual preferences of senders appear to dominate, despite the fact that altruism (and other considerations such as the nature of the relationship and identity management) are both theoretically and self-reportedly present.  The preference-salience model serves to explain why this may be so.  We posit that people select an item to share based on a combination of their preference for it and what is salient at the moment and show that this information can help to predict sharing decisions.

Our work also points to the need for learning more about the motivations for directed sharing and how they might be modeled. Future work addressing how preferences, salience of items and recipients, and other social forces impact sharing decisions will shed more light on people's sharing processes and lead to diffusion models well-suited for online social networks.

\section{Acknowledgments}
This work was supported by the National Science Foundation under grants IIS 0845351, 0910664, and 1422484.

\bibliographystyle{abbrv}
%\bibliography{recco}  

\begin{thebibliography}{10}

\bibitem{sharethis14}
K.~Abrahamson.
\newblock Pinterest surpasses email for sharing online and beats {F}acebook
  growth in 2013.
\newblock {\em Retrieved from
  \url{http://www.sharethis.com/blog/2014/01/16/pinterest-surpasses-email-sharing-online-beats-facebook-growth-2013}},
  January 2014.

\bibitem{aral12-susceptible}
S.~Aral and D.~Walker.
\newblock Identifying influential and susceptible members of social networks.
\newblock {\em Science}, 2012.

\bibitem{bakshy12weakties}
E.~Bakshy, I.~Rosenn, C.~Marlow, and L.~Adamic.
\newblock The role of social networks in information diffusion.
\newblock In {\em Proc. WWW}, 2012.

\bibitem{berger14}
A.~Barasch and J.~Berger.
\newblock Broadcasting and narrowcasting: How audience size impacts what people
  share.
\newblock {\em Journal of Marketing Research}, 2014.

\bibitem{bernstein10}
M.~S. Bernstein, A.~Marcus, D.~R. Karger, and R.~C. Miller.
\newblock Enhancing directed content sharing on the web.
\newblock In {\em Proc. CHI}, 2010.

\bibitem{bond12}
R.~M. Bond, C.~J. Fariss, J.~J. Jones, A.~D. Kramer, C.~Marlow, J.~E. Settle,
  and J.~H. Fowler.
\newblock A 61-million-person experiment in social influence and political
  mobilization.
\newblock {\em Nature}, 2012.

\bibitem{brown87}
J.~J. Brown and P.~H. Reingen.
\newblock Social ties and word-of-mouth referral behavior.
\newblock {\em Journal of Consumer Research}, pages 350--362, 1987.

\bibitem{cha09}
M.~Cha, A.~Mislove, and K.~P. Gummadi.
\newblock A measurement-driven analysis of information propagation in the
  flickr social network.
\newblock In {\em Proc. WWW}, 2009.

\bibitem{chung06}
C.~M. Chung and P.~R. Darke.
\newblock The consumer as advocate: self-relevance, culture, and word-of-mouth.
\newblock {\em Marketing Letters}, 17(4):269--279, 2006.

\bibitem{cohen92}
J.~Cohen.
\newblock A power primer.
\newblock {\em Psychological bulletin}, 1992.

\bibitem{dichter66}
E.~Dichter.
\newblock How word-of-mouth advertising works.
\newblock {\em Harvard Business Review}, 44(6):147--160, 1966.

\bibitem{domingos01}
P.~Domingos and M.~Richardson.
\newblock Mining the network value of customers.
\newblock In {\em Proc. KDD}. ACM, 2001.

\bibitem{goel12}
S.~Goel, D.~J. Watts, and D.~G. Goldstein.
\newblock The structure of online diffusion networks.
\newblock In {\em Proc. ACM Electron. Commerce}, EC '12, 2012.

\bibitem{gruhl04}
D.~Gruhl, R.~Guha, D.~Liben-Nowell, and A.~Tomkins.
\newblock Information diffusion through blogspace.
\newblock In {\em Proc. WWW}, 2004.

\bibitem{hennig-thurau03}
T.~Hennig-Thurau and G.~Walsh.
\newblock Electronic word-of-mouth: Motives for and consequences of reading
  customer articulations on the internet.
\newblock {\em Int. J. Electron. Commerce}, 8(2), Dec. 2003.

\bibitem{ho10}
J.~Y. Ho and M.~Dempsey.
\newblock Viral marketing: Motivations to forward online content.
\newblock {\em J. of Business Research}, 2010.

\bibitem{netflix14}
C.~Johnson.
\newblock Got any good recommendations?
\newblock {\em Retrieved from
  \url{http://blog.netflix.com/2014/09/got-any-good-recommendations.html}},
  September 2014.

\bibitem{kempe03diffusioninfluence}
D.~Kempe, J.~Kleinberg, and {\'E}.~Tardos.
\newblock Maximizing the spread of influence through a social network.
\newblock In {\em Proc. KDD}, 2003.

\bibitem{krauss91}
R.~M. Krauss and S.~R. Fussell.
\newblock Perspective-taking in communication: Representations of others'
  knowledge in reference.
\newblock {\em Social Cognition}, 9(1), 1991.

\bibitem{krishnan08}
V.~Krishnan, P.~K. Narayanashetty, M.~Nathan, R.~T. Davies, and J.~A. Konstan.
\newblock Who predicts better?: results from an online study comparing humans
  and an online recommender system.
\newblock In {\em Proc. RecSys}, 2008.

\bibitem{kulkarni13}
C.~Kulkarni and E.~Chi.
\newblock All the news that's fit to read: a study of social annotations for
  news reading.
\newblock In {\em Proc. CHI}, 2013.

\bibitem{lagnier13probdiffusion}
C.~Lagnier, L.~Denoyer, E.~Gaussier, and P.~Gallinari.
\newblock Predicting information diffusion in social networks using content and
  user’s profiles.
\newblock In {\em Advances in Information Retrieval}. Springer Berlin
  Heidelberg, 2013.

\bibitem{naaman10}
M.~Naaman, J.~Boase, and C.-H. Lai.
\newblock Is it really about me? {M}essage content in social awareness streams.
\newblock In {\em Proc. CSCW}, 2010.

\bibitem{socialtwist13}
J.~Neff.
\newblock Email beats social networks for online offer sharing: Study.
\newblock {\em Retrieved from
  \url{http://adage.com/article/digital/socialtwist-sharing-e-mail-facebook-twitter/244397/}},
  October 2013.

\bibitem{sarwar01}
B.~Sarwar, G.~Karypis, J.~Konstan, and J.~Riedl.
\newblock Item-based collaborative filtering recommendation algorithms.
\newblock In {\em Proc. WWW}, 2001.

\bibitem{sharma11}
A.~Sharma and D.~Cosley.
\newblock Network-centric recommendation: Personalization with and in social
  networks.
\newblock In {\em Proc. IEEE SocialCom}, 2011.

\bibitem{sharma13}
A.~Sharma and D.~Cosley.
\newblock Do social explanations work? {S}tudying and modeling the effects of
  social explanations in recommender systems.
\newblock In {\em Proc. WWW}, 2013.

\bibitem{sharma-icwsm13}
A.~Sharma, M.~Gemici, and D.~Cosley.
\newblock Friends, strangers, and the value of ego networks for recommendation.
\newblock In {\em Proc. ICWSM}, 2013.

\bibitem{susurvey09}
X.~Su and T.~M. Khoshgoftaar.
\newblock A survey of collaborative filtering techniques.
\newblock {\em Adv. in Artif. Intell.}, 2009.

\bibitem{sundaram98}
D.~S. Sundaram, K.~Mitra, and C.~Webster.
\newblock Word-of-mouth communications: a motivational analysis.
\newblock {\em Advances in consumer research}, 1998.

\bibitem{taylor12marketing}
D.~G. Taylor, D.~Strutton, and K.~Thompson.
\newblock Self-enhancement as a motivation for sharing online advertising.
\newblock {\em Journal of Interactive Advertising}, 2012.

\bibitem{taylor13}
S.~J. Taylor, E.~Bakshy, and S.~Aral.
\newblock Selection effects in online sharing: Consequences for peer adoption.
\newblock In {\em Proc. ACM Electron. Commerce}, EC '13, 2013.

\bibitem{wang13}
B.~Wang, C.~Wang, J.~Bu, C.~Chen, W.~V. Zhang, D.~Cai, and X.~He.
\newblock Whom to mention: Expand the diffusion of tweets by @ recommendation
  on micro-blogging systems.
\newblock In {\em Proc. WWW}, 2013.

\bibitem{wasko05}
M.~M. Wasko and S.~Faraj.
\newblock Why should i share? examining social capital and knowledge
  contribution in electronic networks of practice.
\newblock {\em MIS Quarterly}, 2005.

\bibitem{watts02simplecascades}
D.~J. Watts.
\newblock A simple model of global cascades on random networks.
\newblock {\em Proc. PNAS}, 99(9), 2002.

\bibitem{zhao09}
D.~Zhao and M.~B. Rosson.
\newblock How and why people twitter: The role that micro-blogging plays in
  informal communication at work.
\newblock In {\em Proc. GROUP}, 2009.

\end{thebibliography}

\appendix
\subsection{Effects of people and items on observed ratings}
In earlier sections, we presented t-tests for comparing ratings by participants for ease of exposition. However, two empirical observations for our data violate the assumptions for a t-test. First, the distribution of participants' ratings is not normal. Second, ratings by the same user, or for the same item, may not be independent of each other. Thus, we now present a linear mixed-effects analysis to account for non-normality and interdependence in the data.

Mixed-effects analysis (or hierarchical regression) accounts for the interdependence in data by identifying the fixed and random effects on the dependent variable (in our case, people's ratings). We can encode the dependence between ratings by the same user or for the same item as random effects due to user and item, and consider our specific experimental manipulation as the fixed effect. Being a form of regression, the specific assumptions made are that the residual errors have expectation zero, are independent, and have equal variances.

\subsubsection{Senders' ratings for shared and non-shared movies}
When comparing ratings given by senders for shared and non-shared movies as in Table~\ref{tab-ownpref_sig}, whether a movie was shared or not can be considered as a fixed effect on the rating. The sender and the movie are random effects on the rating, which leads us to the following model:
$$ rating \sim shared\_or\_not + (1|participant) + (1|movie) + \epsilon$$
where $\epsilon$ denotes the random error. Using this formulation, we compare this model against a null model which does not incorporate the sharing variable.
$$ rating \sim (1|participant) + (1|movie) + \epsilon$$
We analyze the significance of whether a movie was shared or not by comparing the likelihood of the observed data given our model and the null model. Table~\ref{tab-lme-sharedornot} shows the results, using \textit{lme4} in R (\url{http://CRAN.R-project.org/package=lme4}), including by-participant and by-movie random slopes for the effect of sharing. As before, shared movies are rated significantly higher than non-shared ones in all three groups.

\begin{table}[tbh]
\center
\begin{tabular}{llll}
\tabhead{Group} & $N$ & $\chi^2(1)$  & $p$-value \\ 
\toprule[0.10em]
\bothshown & $609$ &  $6.6$ & $<0.01 $\\
\ownshown & $179$ & $ 5.4$ & $0.02$\\
\othershown & $178$ & $13.6$ & $<0.001$  \\
\midrule
\allcond & $966$ & $23.5$ & $<0.001$  \\
\end{tabular}
\caption{Significance tests using a linear mixed-effects analysis for comparing senders' ratings of shared and non-shared movies. Across all groups, shared movies are rated significantly higher.}
\label{tab-lme-sharedornot}
\end{table}

\begin{table}[tbh]
\center
\begin{tabular}{llll}
\tabhead{Group} & $N$ & $\chi^2(1)$  & $p$-value \\ 
\toprule[0.10em]
\bothshown & $200$ &  $0.63$ & $0.4 $\\
\ownshown & $151$ & $5.1$ & $0.02$\\

\midrule
\textit{Combined} & $351$ & $1.4$ & $0.2$  \\
\end{tabular}
\caption{Significance tests using a linear mixed-effects analysis for comparing recipients' ratings for shared and recommended movies. As we saw before, ratings for shared items are significantly higher than algorithmic recommendations only when senders shared from  movies recommended for the recipient.}
\label{tab-lme-recipient}
\end{table}

\subsubsection{Recipients' ratings for shared and recommended movies}
We can use a similar analysis for comparing recipients' rating for shared and recommended movies. Our null model would be the same as before.
$$ rating \sim (1|participant) + (1|movie) + \epsilon$$
The full model would include information about whether the movie was shared or recommended.
$$ rating \sim shared\_or\_rec + (1|participant) + (1|movie) + \epsilon$$ 
Table~\ref{tab-lme-recipient} shows the comparison between the two models above. The difference between the two models is significant for \ownshown\, but not for \bothshown, similar to our earlier statistical analysis using t-tests. 

\end{document}